\documentclass[runningheads]{svjour2}
\smartqed  
\usepackage{graphicx}
\usepackage{amsmath}
\usepackage{amssymb}

\journalname{Theoretical and Mathematical Physics}

\begin{document}

\title{A perturbative approach for the crystal chains with self-consistent stochastic reservoirs}
\titlerunning{Crystal chains with stochastic reservoirs} 
\author{Ricardo Falcao   \and
        Ant\^onio Francisco Neto \and
        Emmanuel Pereira
}
\institute{Ricardo Falcao \at Departamento de F\'{\i}sica-ICEx,
UFMG, CP 702,
30.161-970 Belo Horizonte MG, Brazil. \\
              \email{rfalcao@fisica.ufmg.br}           
           \and
Ant\^onio Francisco Neto \at N\'ucleo de F\'{\i}sica, Campus
Universit\'ario Prof. Alberto Carvalho-UFS 49500-000 Itabaiana,
SE, Brazil. \\
\email{afneto@fisica.ufs.br} \and Emmanuel Pereira \at Departamento
de F\'{\i}sica-ICEx, UFMG, CP 702,
30.161-970 Belo Horizonte MG, Brazil. \\
              \email{emmanuel@fisica.ufmg.br}
}
\date{Accepted: 30/10/2007}

\maketitle

\begin{abstract}
We consider the harmonic chain of oscillators with self-consistent
stochastic reservoirs and give a new proof for the finitude of its
thermal conductivity in the steady state. The approach, with
involves an integral representation for the correlations (heat flow)
and a perturbative analysis, is quite general and extendable to the
study of anharmonic systems.

\end{abstract}
\maketitle

In nonequilibrium statistical physics, the analytical derivation of
the macroscopic laws of thermodynamics for microscopic models of
interacting particles is a challenge to theorists. In particular,
even in 1D context, it is still unknown the rigorous derivation of
Fourier's law of heat conduction which states that the heat flow is
proportional to the gradient of the temperature \cite{BLR}. Many
works, almost all of them by means of computer simulations
\cite{LLP} and with conflicting results have been devoted to the
theme since the pioneering rigorous study on the harmonic chain of
oscillators with thermal baths at the boundaries \cite{RLL}, a model
that does not obey the Fourier's law. Recently, in the scenario of
analytical studies, the harmonic chain of oscillators has been
revisited, but in the case of each site coupled to a stochastic
reservoir: it is proved that the Fourier's law holds in such case
\cite{BLL}; of course, if we turn off the coupling with the inner
reservoirs, the heat conductivity diverges and the Fourier's law
does not hold anymore, as previously shown in \cite{RLL}. As the
procedure presented in \cite{BLL} was completely dependent on the
linearity of the dynamics (i.e., on the harmonicity of the
potential), some of the present authors have proposed a different
approach to study the chain of oscillators with a reservoir at each
site in order to analyze the problem of anharmonic potentials and
infer their behavior with thermal baths at the boundaries only
\cite{PF1,PF2,FLP}. The proposed approach involves an integral
representation for the correlations (related to the heat flow),
whose analysis is carried out by means of a perturbative
computation.

In the present letter, as a rigorous support for our previous
theoretical works (which involve, as said, anharmonic interactions),
we turn to the simpler case of the harmonic chain with a reservoir
at each site and give a new proof for the finitude of the thermal
conductivity (that shall be extendable to the anharmonic case),
using our approach and a perturbative analysis: the expression for
the conductivity presented here coincide with that one described in
\cite{BLL}.

Let us introduce the model. We consider a lattice system with
unbounded scalar variables in a space $\Lambda \subset \mathbb{Z}$
and with stochastic heat bath at each site. Precisely, we take a
system of $N$ oscillators with the Hamiltonian
\begin{equation}
H(p,q)=\sum_{j=1}^N\frac{1}{2}\left (
p_j^2+Mq_j^2\right)+\frac{1}{2}\sum_{j\neq l=1}^N q_jJ_{jl}q_l,
\label{hamiltonian}
\end{equation}
where $J_{lj}=J_{jl}$ and $J_{lj}=f(|l-j|)$. In the study of
harmonic chains it is usually taken $j$ and $l$ nearest neighbors
and $J_{j,j+1}=J$ for any $j$. In this case, the interparticle
interations may be also written as (assuming Dirichlet boundary
conditions $q_0=0=q_{N+1}$)
\begin{equation}
\sum_{j=1}^NV(q_j-q_{j+1})~~,~~V=\frac{J^2}{2}q^2,
\end{equation}
after adjustments in $M$, the coefficient of $q_j^2$. We consider
the Langevin time evolution given by the stochastic differential
equations
\begin{eqnarray}
dq_j&=&p_jdt, \nonumber\\
dp_j&=&-\frac{\partial H}{\partial q_j}dt-\xi
p_jdt+\gamma_j^{1/2}dB_j,~~j=1,2,\cdots N,
\end{eqnarray}
where $B_j$ are independent Wiener processes (i.e., $\eta_j=dB_j/dt
$ are independent white noises) $\xi$ is the coupling between the
site $j$ and its heat bath; $\gamma_j\equiv 2\xi T_j$, where $T_j$
is the temperature of the j-th reservoir. As usual, we define the
energy of a single oscillator as
\begin{equation}
H_j(q,p)=\frac{1}{2}p_j^2+q_j^2+U(q_j)+\frac{1}{2}\sum_{l\neq j}
V(q_j-q_l),
\end{equation}
where the expressions for $U$ and $V$ follow from eq.
(\ref{hamiltonian}) and from $\sum_j H_j=H$. Then for the energy
current we have
\begin{equation}
\left <\frac{dH_j}{dt} \right >=\left < R_j \right >-\left
<\mathcal{F}_{j\leftarrow}-\mathcal{F}_{j\rightarrow} \right>,
\end{equation}
where $\left<\cdot \right>$ means the expectation with respect to
the noise distribution; $R_j$ denotes the energy flux between the
j-th reservoir and the j-th site
\begin{equation}
\left < R_j\right>=\xi\left ( T_j-\left <p_j^2\right > \right );
\end{equation}
and the energy current in the chain is
\begin{eqnarray}
\mathcal{F}_{j\rightarrow}=\sum_{l>j}\nabla
V(q_j-q_l)\frac{p_j+p_l}{2}, \nonumber \\
\mathcal{F}_{j\leftarrow}=\sum_{l<j}\nabla
V(q_l-q_j)\frac{p_l+p_j}{2}. \label{flux}
\end{eqnarray}
As well known, the stationary state is characterized by $\left <
dH_i/dt \right >=0$; and, for physical reasons, it is interesting to
consider the ``self-consistent condition"  given by $\left <
R_j\right>=0$ in the steady state. For the linear dynamics, the
existence and convergence to the stationary state as
$t\rightarrow\infty$ are old solved problems, see e.g. \cite{SZ}.

Lets us state our main theorem.

\begin{theorem}\label{teo1}
For the harmonic chain of oscillators with reservoirs at each site
(1-3), in the case of nearest neighbor interactions, i.e.
$J_{jl}=J(\delta_{l,j+1}+\delta_{l,j-1})$, with $J<J_0$ for some
small $J_0$, the Fourier's law holds
\begin{equation}
\mathcal{F}=\lim_{t\to \infty} \left <\mathcal{F}_{j\rightarrow}
\right>=-\frac{\chi}{N-1}(T_N-T_1),
\end{equation}
with the ``self consistent" condition $\lim_{t\to\infty}\left <R_j
\right
>=0$ for the inner sites $j$, and with the heat conductivity $\chi$
given by
\begin{equation}
\chi=\frac{J^2}{2\xi M}+\mathcal{O}(J^3).
\end{equation}
\end{theorem}
\begin{remark} A theorem establishing the finitude of the thermal
conductivity for the considered model has been already presented in
\cite{BLL}, as already said, and with the nearest neighbor
interparticle potential not necessarily small. Our aim in this
letter, as emphasized, is not to give a second proof but to present
a more general approach which shall extend to the anharmonic chains.
\end{remark}
\begin{remark} Following the steps to be presented ahead, we may also study the
heat flow for weak interparticle interactions beyond nearest
neighbor: e.g. for $\sup_{l}\sum_j J_{lj}\leq J_0$.
\end{remark}
Now we describe our approach. It is usefull to introduce the
phase-space vector $\phi=(q,p)$ and write the dynamics as
\begin{equation}
\dot{\phi}=-A\phi+\sigma\eta, \label{dynamics}
\end{equation}
where $A=A^0+\mathcal{J}$ and $\sigma$ are $2N\times 2N$ matrices
\begin{eqnarray}
A^0=\left ( \begin{array}{cc} 0 & - I \\
\mathcal{M} & \Gamma  \end{array} \right ),\mathcal{J}=\left (
\begin{array}{cc} 0 & 0 \\
\mathbb{J} & 0 \end{array} \right ),\sigma=\left (\begin{array}{cc} 0 & 0 \\
0 & \sqrt{2\Gamma\mathcal{T}} \end{array}\right ),
\end{eqnarray}
where $I$ is the unit $N\times N$ matrix, $\mathbb{J}$ is the
$N\times N$ matrix for the interparticle interactions, and
$\mathcal{M}$, $\Gamma$ and $\mathcal{T}$ are the diagonal $N\times
N$ matrices: $M_{jl}=M\delta_{jl}$, $\Gamma_{jl}=\xi\delta_{jl}$,
$\mathcal{T}=T_j\delta_{jl}$. And $\eta$ are independent white
noises. In what follows we will use the index notation: $i$ for
index values in the set $[N+1,N+2,\cdots,2N]$; $j$ for values in the
set $[1,2,\cdots,N]$ and $k$ for values in $[1,2,\cdots ,2N]$. We
will also omit obvious sums over repeated indices.

Our strategy is first to consider (\ref{dynamics}) above with
$\mathcal{J}\equiv 0$, i.e., a system with isolated sites (without
interactions among them). In a second step we introduce the
interparticle interaction by using the Girsanov theorem, and then we
calculate the heat flow in a perturbative computation.

We have

\begin{lemma}
The solution of (\ref{dynamics}) with $\mathcal{J}\equiv 0 $ is the
Ornstein-Uhlenbeck Gaussian process
\begin{equation}
\phi(t)=e^{-tA^0}\phi(0)+\int_0^t dse^{-(t-s)A^0}\sigma\eta(s),
\end{equation}
where, for the simple case of $\phi(0)=0$, the covariance of the
process evolves as
\begin{eqnarray}
\left <\phi(t)\phi(s)\right >_0\equiv \mathcal{C}(t,s)=\left \{
\begin{array}{c} e^{-(t-s)A^0}\mathcal{C}(s,s), ~~t\geq s,\label{covariance1} \\
\mathcal{C}(t,t)e^{-(s-t)A^{0^\top}}, ~~t\leq s,\end{array}\right . \\
\mathcal{C}(t,t)=\int_0^tdse^{-sA^0}\sigma^2e^{-sA^{0^\top}}.
\label{covariance}
\end{eqnarray}
\end{lemma}
\begin{proof} It is a simple exercise of stochastic differential equations
(see e.g. \cite{Ok}, p.74 )$\qed$.
\end{proof}
In the simple case of $\mathcal{J}\equiv0$, as $t\to \infty$ we have
a convergence to the Gaussian distribution with covariance
\begin{eqnarray}
C=\int_0^{\infty}ds e^{-sA^0}\sigma^2e^{-sA^{0^\top}}=\left
(\begin{array}{cc}\frac{\mathcal{T}}{\mathcal{M}} &0 \\
0&\mathcal{T}\end{array} \right),\label{ecovariance}
\end{eqnarray}
where as said , $\mathcal{T}$ is a diagonal $N\times N$ matrix with
$\mathcal{T}_{ij}=T_i\delta_{ij}$ (see e.g. \cite{SZ}).

The solution of (\ref{dynamics}) with the interparticle potential
will be derived from the particular case of $\mathcal{J}\equiv 0$
by using the Girsanov theorem which establishes a measure $\rho$
for the complete process in terms of the measure
$\mu_{\mathcal{C}}$ obtained for $\mathcal{J}\equiv0$. Precisely,
for the two-point function we have

\begin{lemma}The two-point functions for the complete process
(\ref{dynamics}) can be written as
\begin{equation}
\left <\varphi_u(t_1)\varphi_m
(t_2)\right>=\int\phi_u(t_1)\phi_m(t_2)Z(t)d\mu_{\mathcal{C}}/{\rm
norm.}, ~~t_1,t_2<t,\label{tpf}
\end{equation}
where $\phi$ is the solution (given by Lemma 1) of the process with
$\mathcal{J}\equiv 0$, and $\varphi$ is the solution for the
complete process (\ref{dynamics}). $\mathcal{C}$ is given by
(\ref{covariance1},\ref{covariance}), and the corrective factor is
\begin{eqnarray}
Z(t)=\exp\left (\int_0^t u\cdot dB-\frac{1}{2}\int_0^tu^2 ds\right ), \nonumber\\
\gamma_i^{1/2}u_i=-J_{i-N,j}\phi_j .
\end{eqnarray}
\end{lemma}
\begin{proof} For the harmonic potential, the process is an It\^{o}
diffusion, and so, the proof is also direct: see e.g. theorem
$8.6.8$ in \cite{Ok}. $\qed$
\end{proof}
\begin{remark} In the case of the nonlinear (anharmonic) dynamical
system problem we may introduce the anharmonic interactions (on-site
and/or interparticle potentials) still by using the Girsanov
theorem: e.g., for bounded potentials or initial processes with
continuous realization, the use of Novikov condition makes the
procedure straightforward; see an example of other manipulations in
a similar use of the Girsanov theorem \cite{D}.
\end{remark}
Turning to the $Z(t)$ expression above, we have
\begin{eqnarray}
u_idB_i&=&\gamma_i^{-1/2}u_i\gamma_i^{1/2}dB_i \nonumber \\
&=&\gamma_i^{-1/2}u_i\left (d\phi_i+A_{ik}^0\phi_kdt \right )
\nonumber \\
&=&-\gamma_i^{-1/2}\mathcal{J}_{ij}\phi_j\left
(d\phi_i+A_{ik}^0\phi_kdt\right).
\end{eqnarray}
Using the It\^{o} formula we get
\begin{eqnarray}
&-\gamma_i^{-1}\mathcal{J}_{ij}\phi_j
d\phi_i=-dF-\gamma_i^{-1}\phi_i\mathcal{J}_{ij}A_{jk}^0\phi_k
dt,& \label{udb} \\
&F(\phi)=\gamma_i^{-1}\phi_i\mathcal{J}_{ij}\phi_j .& \nonumber
\end{eqnarray}
And so,
\begin{eqnarray}
Z(t)&=&\exp\left(\int_0^tu\cdot dB -\frac{1}{2}\int_0^t u^2 ds
\right)
\label{zt} \\
&=&\exp\left[-F(\phi(t))+F(\phi(0)) \right]\exp\left[-\int_0^t
W(\phi(s))ds \right ] ,\nonumber \\
W(\phi(s))&=&\gamma_i^{-1}\phi_i(s)\mathcal{J}_{ij}A_{jk}^0\phi_k(s)+\phi_k(s)A_{ki}^{0^\top}\gamma_i^{-1}\mathcal{J}_{ij}\phi_j+\label{w1}
\\
&+&\frac{1}{2}\phi_{j'}(s)\mathcal{J}_{j'i}\gamma_i^{-1}\mathcal{J}_{ij}\phi_j(s).\nonumber
\end{eqnarray}
To analyze the heat flow in the steady state (related to
$\lim_{t\to\infty}\left<\varphi_u(t)\varphi_v(t) \right>$), we
note that we may write
\begin{eqnarray}
\exp\left (-tA^0 \right )=e^{-t\frac{\xi}{2}}\cosh(t\rho)\left \{
\left (\begin{array}{cc}I & 0 \\0&I \end{array}\right )
+\frac{\tanh(t\rho)}{\rho}\left (
\begin{array}{cc}
\frac{\xi}{2} & I \\
-\mathcal{M} & -\frac{\xi}{2}
\end{array} \right ) \right \},\label{exp}
\end{eqnarray}
$\rho=\left(\left(\frac{\xi}{2}\right)^2-M\right)^{1/2}$. It may
be shown by e.g., diagonalizing $A^0$ (or see the appendix of
\cite{BLL}). We also note that
\begin{equation}
\mathcal{C}(t,s)=\exp\left(-(t-s)A^0
\right)C+\mathcal{O}(\exp\left(-(t+s)\xi/2
\right)),\label{scovariance}
\end{equation}
where the effects in the correlation function formula of the
second term on the right-hand side of the equation above vanishes
in the limit $t\to\infty$; $C$ is the covariance
(\ref{ecovariance}).

Now, to compute the two-point correlation function (\ref{tpf}) (and
so, the heat flow), we expand the exponential which gives $Z(t)$
above (\ref{zt}-\ref{w1}) in a power series:
$\exp[X]=\sum_{n=0}^{\infty}X^n/n!$, and calculate the connected
Feynman graphs for (\ref{tpf})(due to the normalization factor we
stay with the connected graphs only) using the Wick theorem. As we
have quadratic terms in $\phi$ in the exponent (see $F$ and $W$
above), there is no countable problem with the series; roughly,
\begin{equation}
\left |\int\frac{\phi^{2n}}{n!}d\mu_{\mathcal{C}} \right
|=\left|\frac{1}{n!}\sum_{k=1}^{(2n-1)!!}\underbrace{(\mathcal{C}\mathcal{C}\cdots\mathcal{C})}_n\right|\leq\frac{(2n)!!}{n!}\left
|\mathcal{C}\cdots\mathcal{C} \right|=2^n\left
|\mathcal{C}\cdots\mathcal{C} \right |.
\end{equation}
In short, to control the expansion we only need a bound for the $n$
convolutions of the covariance $\mathcal{C}$ such as $c^nJ^n$, where
$c$ is some constant. Thus at least for small $J$, this simple
analysis will give us the convergence of the perturbative series. To
get such bound, we use the formulas (\ref{exp}-\ref{scovariance})
for $\mathcal{C}$, the lemma below and that
\begin{eqnarray*}
&||C||\leq c_1(1-e^{-2\alpha t}),&\\
&||e^{-tA^0}||\leq c_2e^{-\alpha t},&\\
&||A^0||\leq c_3,&\\
\end{eqnarray*}
where $\alpha={\rm min}\{\xi/2,M/\xi\}$ ($||\bullet||$ means a
matrix bound on $M_{2N\times2N}(\mathbb{R})$).

\begin{lemma}\label{lemma3}
Let $I_t$ be
\begin{equation}
I_t=\int_0^t e^{-\alpha|t-s_1|}e^{-\alpha|s_1-s_2|}\cdots
e^{-\alpha|t-s_n|}ds_1\cdots ds_n, ~~\alpha >0,
\end{equation}
then, $\lim_{t\to\infty}I_t \leq (c_\alpha)^n$, where $c_{\alpha}$
does not depend on $n$.
\end{lemma}
\begin{proof} For $f(x)=e^{-\alpha|x|}$, we have
\begin{equation}
\tilde{f}(p)\equiv\frac{1}{\sqrt{2
\pi}}\int_{-\infty}^{\infty}e^{-ipx}e^{-\alpha|x|}dx=\frac{1}{\sqrt{2\pi}}\frac{2\alpha}{\alpha^2+p^2}.
\end{equation}
And
\begin{eqnarray}
I_t&\leq&\frac{1}{2}\int_{-\infty}^{\infty}e^{-\alpha|t-s_1|}e^{-\alpha|s_1-s_2|}\cdots
e^{-\alpha|t-s_n|}ds_1\cdots ds_n \nonumber \\
&=&\frac{1}{2}\int_{-\infty}^{\infty}f(t-s_1)f(s_1-s_2)\cdots
f(s_n-t)ds_1\cdots ds_n\nonumber \\
&=&\frac{1}{2}\underbrace{f\ast f\ast\cdots
\ast f}_{n+1}(0) \nonumber \\
&=&\frac{1}{2}(2\pi)^{(n-1)/2}\int_{-\infty}^{\infty}(\tilde{f}(p))^{n+1}dp,
\end{eqnarray} where $*$ means the convolution, and we have used
Parseval's theorem in the last equality above. Hence,
\begin{eqnarray}
\lim_{t\to\infty}I_t\leq
\frac{1}{2}(2\pi)^{(n-1)/2}\int_{-\infty}^{\infty}\frac{1}{(2\pi)^{(n+1)/2}}\left
(\frac{2\alpha}{\alpha^2+p^2}\right)^{n+1}dp\leq (c_{\alpha})^n.
\end{eqnarray}
\end{proof}
Thus, using the formulas (\ref{exp}-\ref{scovariance}) for the
covariance and the lemma \ref{lemma3} above, we may bound the terms
order $J^2$ and up in the two point function (\ref{tpf}) by
\begin{equation}
\sum_{n=2}^{\infty}(c'J)^n\leq c''J^2, \label{cota}
\end{equation}
where $c'$ and $c''$ are some constants (i.e., they do not depend on
$n$).

To obtain the result of theorem \ref{teo1} we still need to
calculate, in detail, the two-point correlation function up to first
order in $J$. Carrying out the computation (simple Gaussian
integrations), for $\left
<\varphi_u\varphi_v\right>\equiv\lim_{t\to\infty}\left<\varphi_u(t)\varphi_v(t)
\right>$, we obtain
\begin{eqnarray}
\left<\varphi_u\varphi_v\right >= \left
\{\begin{array}{l}\frac{1}{2\xi M}\left
[\mathcal{J}_{v+N,u-N}T_{u-N}-\mathcal{J}_{u,v}Tv\right
],\textrm{for} \left [\begin{array}{l}u\in [N+1,\cdots,2N], \\ v\in[1,\cdots,N],\end{array}\right . \\
 T_{u-N}\delta_{u,v},~~~~~~~~~~~~~~~~~~~~~~~~~~~~~~ \textrm{for}~u,v\in [N+1\cdots,2N].
\end{array}\right.\label{result}
\end{eqnarray}
Hence, from (\ref{flux})
\begin{equation}
\mathcal{F}_{j\to}=\sum_{r>j}\mathcal{J}_{j+N,r}\left
(\varphi_j-\varphi_r
\right)\frac{(\varphi_{j+N}+\varphi_{r+N})}{2},~~r \in[1,\cdots
N],\label{flux1}
\end{equation}
and we have (for $\left <\varphi_u\varphi_v \right>$ up to first
order in $J)$
\begin{equation}
\left <\mathcal{F}_{j\to}\right
>=\sum_{r>j}\frac{\left(\mathcal{J}_{j+N,r}\right)^2}{2\xi
M}\left (T_{r}-T_j\right),
\end{equation}
or, for nearest neighbor interactions,
\begin{equation}
\mathcal{F}_{j\to j+1}=\left <\mathcal{F}_{j\to}\right
>=\frac{\left(\mathcal{J}_{j+N,j+1}\right)^2}{2\xi
M}\left (T_{j+1}-T_j\right).
\end{equation}
The first order perturbation computation (\ref{result}) above still
gives us $$ \lim_{t\to \infty}\left < R_{j}(t)\right >=0.$$

Thus, the steady state condition $\left < dH_i/dt\right>=0$ leads to
\begin{equation}
\mathcal{F}_{1\to 2}=\mathcal{F}_{2\to
3}=\cdots=\mathcal{F}_{N-1\to N}\equiv\mathcal{F}.
\end{equation}
And so, for the simpler case of the same interactions between any
two nearest neighbors sites, i.e., $J=\mathcal{J}_{j+N,j+1}$ for any
$j$, it follows that (for $\left <\varphi_u\varphi_v \right
>$ up to first order in $J$), we have
\begin{equation}
\mathcal{F}_{\to}=\chi\frac{T_N-T_1}{N-1},~~~\chi=\frac{J^2}{2\xi
M}.\label{flux2}
\end{equation}
From (\ref{flux1}) and (\ref{cota}), it is easy to see that
considering the remaining terms (higher order in $J$), we get as
$J\times c J^2=cJ^3$ in $\chi$ (the temperature terms $T_{j+1}-T_j$
may be extract from the perturbative analysis following the products
of $C$, the covariance (\ref{ecovariance}) in $\mathcal{C}$ which
appears in the Gaussian interaction, and the terms $\gamma_i$ coming
from the expansion of $Z(t)$). In short, theorem \ref{teo1} holds.

\begin{remark} In the case of anharmonic terms in $W$ and $F$
(\ref{udb}-\ref{w1}), a naive expansion fails, but we expect to
develop a perturbative approach by using some cluster expansion as
well known in field theory and statistical mechanics. For the
simpler case of nonconservative nonlinear stochastic dynamical model
(describing a system in contact with thermal reservoirs at the same
temperature, and so going to equilibrium), a convergent cluster
expansion is presented in \cite{D}, and the decay of the four-point
function is investigated in \cite{SBFP} and \cite{FOPS}: the
complete and rigorous result \cite{FOPS} adds only small corrections
to the first order perturbative calculation \cite{SBFP}.
\end{remark}
\begin{acknowledgements}
Work partially support by CNPq and CAPES.
\end{acknowledgements}


\begin{thebibliography}{3}

\bibitem{BLR} Bonetto, F., Lebowtiz,  J. L.,  Rey-Bellet,  L.:  Fourier's Law: a challenge to
theorists. In Fokas, A.,  Grigoryan, A.,  Kibble, T., Zegarlinski,
B. (ed.)  Mathematical Physics, London, 2000,  128-150, Imperial
College Press, (2000)

\bibitem{LLP} Lepri, S., Livi, R., Politi, A.: Thermal conduction in classical low-dimensional lattices.  Phys. Rep. {\bf 377}, 1--80 (2003)

\bibitem{RLL} Rieder, Z., Lebowtiz, J. L., Lieb,  E.: Properties of a harmonic crystal in a stationary nonequilibrium state.  J. Math. Phys. {\bf 8}, 1073--1078 (1967)

\bibitem{BLL} Bonetto, F., Lebowitz,  J. L.,  Lukkarinen, J.: Fourier's law for a harmonic crystal wit self-consistent stochastic reservoirs.  J. Stat.Phys {\bf 116}, 783--813 (2004)

\bibitem{PF1} Pereira, E., Falcao, R.: Nonequilibrium statistical mechanics of anharmonic crystal with self-consistent stochastic reservoirs.  Phys. Rev.E {\bf 70}, 046105-1--046105-5 (2004)

\bibitem{PF2} Pereira, E., Falcao, R.: Normal heat conduction in a chain with a weak interparticle anharmonic potential.  Phys. Rev. Lett. {\bf 96}, 100601-1--100601-4 (2006)

\bibitem{FLP} Francisco Neto, A.,  Lemos, H. C. F., Pereira,  E.: Heat conduction in a weakly anharmonic chain: an analytical approach.  J. Phys. A {\bf 39}, 9399--9410 (2006)

\bibitem{SZ} Snyders, J.,  Zakai, M.: On nonnegative solutions of the equation $AD+DA'=-C^\ast$. J.Appl. Math. {\bf 18}, 704--714 (1970)

\bibitem{Ok} {\O}ksendal, B.: Stochastic Differential Equations: An Introduction with Applications. Springer-Verlag, Berlin
(2003)

\bibitem{D}  Dimock, J.: A cluster expansion for stochastic lattice fields,   J. Stat. Phys. {\bf 58},  1181--1207  (1990)

\bibitem{FOPS} Faria da Veiga, P. A.,   O'Carroll, M.,  Pereira, E., Schor, R.: Spectral analysis of weakly coupled stochastic lattice Ginzburg-Landau models,  Commun. Math. Phys. {\bf 220}, 377-402 (2001)

\bibitem{SBFP} Schor, R.,  Barata, J. C. A.,  Faria da Veiga, P. A.,  Pereira, E.: Spectral properties of weakly coupled Landau-Ginzburg stochastic models,  Phys. Rev. E {\bf 59}, 2689--2694 (1999)

\end{thebibliography}
\end{document}